\documentclass[twocolumn,showpacs,preprintnumbers,amsmath,amssymb]{revtex4-1}
\usepackage{amsmath,amsfonts,amssymb}
\usepackage{graphicx}

\begin{document}
\title{Gravitating tensor monopole in a Lorentz-violating field theory}
\author{Xin-zhou Li}
\email{kychz@shnu.edu.cn}
\author{Ping Xi}
\email{xiping@shnu.edu.cn}
\author{Qun Zhang}
\email{1000304240@shnu.edu.cn}
\affiliation{Center for Astrophysics,Shanghai Normal University,100 Guilin Road,Shanghai,200234,China}
\date{\today}

\begin{abstract}
We present a solution of the coupled Einstein and rank-two antisymmetric tensor field equations where Lorentz symmetry is spontaneously broken, and we discuss its observational signatures. Especially, the deflection angles have important qualitative differences between tensor and scalar monopoles. If a monopole were to be detected, it would be discriminated whether or not to correspond to a tensor one. This phenomenon might open up new direction in the search of Lorentz violation with future astrophysical observations.
\end{abstract}
\maketitle

\section{INTRODUCTION}
Lorentz symmetry is cornerstone in the foundation of modern physics. The experimental tests of Lorentz violation are also interested for a decade (see Ref.\cite{11} and references therein). The possibility of Lorentz-violating field theory were intensively studied in the various contexts, including Riemann-Cartan geometry \cite{7}, Riemann-Finsler geometry \cite{8}, string theory \cite{9}, and noncommutative geometry \cite{10}. Especially, a tensor field theory with dynamical Lorentz symmetry violating such that the manifold of equivalent vacua after the violation is not shrinkable to a point may contain monopole solutions \cite{1,2}. There exist monopole solutions in the minimal model coupled to gravity \cite{1,2} for antisymmetric 2-tensor field, in which the far-field approximation and Bogomol'nyi-Prasad-Sommerfield (BPS) limit are used. It is worth noting that above-mentioned solution is not exact one for the metric around a tensor monopole since it is not the solution of the equation of the tensor field. One must still find a solution of the coupled equations of motion valid throughout space.

On the other hand, if the symmetry that is broken is a global symmetry of scalar fields, the gravitational effect of monopole configuration \cite{3} is equivalent to that of a deficit solid angle in the metric, plus that of a negative mass at the origin \cite{4,5}. The properties of scalar monopoles have been investigated in the various space-time \cite{6}. Monopoles could be produced by the phase transition in the early Universe and their existence has important implications in cosmology. It is possible that the monopole still exist as relic object in the Universe today, since isolated topological defect is stable. If a tensor monopole were to be detected, it would offer precious enlightenment on fundamental symmetries in physics. For the scalar case, the internal symmetry is spontaneously broken, and Lorentz symmetry is exact. On the contrary, Lorentz symmetry will be broken by the vacuum solution in the tensor case. However, the signature of tensor monopole is effectively the same as a scalar one in the Seifert's approximation. It is of course not possible to use tensor monopole set-ups to assess the existence of Lorentz violation in this approximation.

In this paper, we propose a novel approach that might relax the above problem and open new possibilities for the detection of Lorentz violation in future astrophysical observations. We show the realistic solution for the coupled system, and discuss its observational signatures. Using standard techniques, we have calculated the light ray propagating in these backgrounds. Note that the deflection angles are dependent upon the "apparent impact parameter". From this point of view, the signature of an antisymmetric tensor monopole can be distinguished from two species of monopoles in the future tests. Moreover, the tensor monopole would provide inestimable insight into the role played by Lorentz symmetry in physics.\\

\section{ The field equations}
We consider the 1+3 dimensional action
\begin{equation}
S=\int{ d^{4}x\sqrt{-g}(\frac{R}{16\pi G}-\mathcal{L}_m)},
\end{equation}
where the gravity part of the action is the usual Einstein-Hilbert action with the gravitational coupling constant $G$ and curvature scalar $R$. $\mathcal{L}_m$ is the Lagrangian of an antisymmetric 2-tensor field which takes on a background expected value \cite{1},
\begin{equation}
\mathcal{L}_m=-\frac{1}{6}F^{\mu\nu\rho}F_{\mu\nu\rho}-\frac{\lambda}{2}(B^{\mu\nu}B_{\mu\nu}-b^2)^2,
\end{equation}
where $B_{\mu\nu}$ is an antisymmetric tensor field and $F_{\mu\nu\rho}=3\partial _{[\mu}B_{\nu\rho]}$ is its associated field strength. Sometime, $B_{\mu\nu}$ is known as the Kalb-Ramond field \cite{13,14}. For the metric, the spherically symmetric ans\"{a}tz in Schwarzschild-like coordinates reads:
\begin{equation}
ds^2=-E(r)dt^2+F(r)dr^2+r^2(d\theta^2+\sin^2\theta d\varphi^2),
\end{equation}
while for the Kalb-Ramond field, we also choose the spherically symmetric ans\"{a}tz \cite{1}:
\begin{eqnarray}
B_{tr}=-B_{rt}=0, B_{\theta\varphi}=-B_{\varphi\theta}=g(r)r^2\sin^2\theta.
\end{eqnarray}
Using Eqs.(1)-(4), the equation of motion for the Kalb-Ramond field can be reduced to
\begin{equation}
\frac{1}{2}(\frac{E'}{E}-\frac{F'}{F})(g'+\frac{2}{r}g)+\frac{\partial}{\partial r}(g'+\frac{2}{r}g)-2\lambda Fg(2g^2-b^2)=0
\end{equation}
where primes denote differentiation with respect to $r$.

As a general feature, a solution with $g(0)=0$ and $g(r)\rightarrow b/\sqrt2$ as $r\rightarrow \infty$ corresponds to a monopole configuration since the vacuum manifold contains a non-contractible two-sphere (i.e., $\pi_2(\mathcal{M}_{vac})=\mathbb{Z}$). Let us mention, by the way, that the equation of motion is analogous between the tensor monopole and $O(3)$ scalar monopole. In the both cases, no exact expression is known, although series method \cite{4} or numerical calculation \cite{5} can be used to approximate it for the $O(3)$ monopole. The energy-momentum tensor of a tensor monopole configuration is
\begin{eqnarray}
T_{t}^{t}&=&-\frac{1}{F}(g'+\frac{2}{r}g)^2-\frac{\lambda}{2}(2g^2-b^2)^2,\\
T_{r}^{r}&=&\frac{1}{F}(g'+\frac{2}{r}g)^2-\frac{\lambda}{2}(2g^2-b^2)^2,\\
T_{\theta}^{\theta}&=&T_{\varphi}^{\varphi}=\frac{1}{F}(g'+\frac{2}{r}g)^2-\frac{\lambda}{2}(2g^2-b^2)^2+4\lambda g^2(2g^2-b^2).\nonumber\\
\end{eqnarray}

Varying the action (1) with respect to the metric fields gives the Einstein equations
\begin{eqnarray}
-\frac{1}{F}(\frac{1}{r^2}-\frac{F'}{Fr})+\frac{1}{r^2}&=&\frac{\epsilon}{2b^2}[\frac{1}{F}(g'+2\frac{g}{r})^2+\frac{\lambda}{2}(2g^2-b^2)^2]\nonumber\\
\\
-\frac{1}{F}(\frac{1}{r^2}+\frac{E'}{Er})+\frac{1}{r^2}&=&\frac{\epsilon}{2b^2}[-\frac{1}{F}(g'+2\frac{g}{r})^2+\frac{\lambda}{2}(2g^2-b^2)^2]\nonumber\\
\end{eqnarray}
where the dimensionless quantity $\epsilon\equiv16\pi G b^2$.

In order to solve the system of equations (5),(9) and (10) uniquely, we have to introduce 6 boundary conditions, which we choose to be
\begin{eqnarray}
&g&(0)=0,F(0)=1,E(0)=e_0,\noindent \nonumber\\
&g&(r)|_{r\rightarrow\infty}=\frac{b}{\sqrt{2}},E(r)r^{-2\epsilon}|_{r\rightarrow\infty}=(2\lambda b^2)^{\epsilon},\noindent \nonumber\\&F&(r)|_{r\rightarrow\infty}=1+\epsilon.
\end{eqnarray}\\

\section{THIN-WALL APPROXIMATION}
We start our discussion with a simplified model for the monopole configuration, just to show the main features of the exact solution in a simple manner. Let us modeling the monopole configuration in the thin-wall limit
\begin{eqnarray}
g= \left\{\begin{array}{ll}
 0&\textrm{if $r<\delta$}\\
\frac{b}{\sqrt{2}}&\textrm{if $r>\delta$}
\end{array}\right.
\end{eqnarray}
where $\delta$ is the core radius. Einstein equations inside the core are solved by a de Sitter metric
\begin{equation}
ds^2=-(1-\frac{\lambda\epsilon b^2r^2}{12})dt^2+\frac{dr^2}{1-\frac{\lambda\epsilon b^2r^2}{12}}+r^2d\Omega^2.
\end{equation}
The exterior solution is given by
\begin{eqnarray}
ds^2=&-&(\sqrt{2\lambda}br)^{2\epsilon}(1-\frac{2GM}{(\sqrt{2\lambda}b)^\epsilon r^{1+\epsilon}})dt^2 \nonumber\\
&+&\frac{1+\epsilon}{1-\frac{2GM}{(\sqrt{2\lambda}b)^\epsilon r^{1+\epsilon}}}dr^2+r^2d\Omega^2,
\end{eqnarray}
where $M$ is an arbitrary constant of integration. Both $\delta$ and $M$ are determined by Eqs.(9) and (10) at the boundary between the interior and exterior region, which correspond to the continuity of the metric. The result is
\begin{eqnarray}
\delta&=&\frac{1}{\sqrt{2\lambda}b}(\frac{1}{1+\epsilon})^\frac{1}{2\epsilon}\\
M&=&-\frac{8\pi b}{\sqrt{2\lambda}}[1-\frac{(1+\epsilon)^{1-\frac{1}{\epsilon}}}{24}](\frac{1}{1+\epsilon})^{\frac{1+\epsilon}{2\epsilon}}
\end{eqnarray}
We argue that it is possible to match an interior de Sitter solution to an exterior tensor monopole solution, but only for $M<0$. This property is consistent with the negative mass of scalar monopole \cite{4}. Furthermore, we have $\frac{1}{\sqrt{2\lambda e}b}\leq\delta\leq\frac{1}{2\sqrt{\lambda}b}$ and $-\frac{23\pi b}{\sqrt{72\lambda}}\leq M\leq-(8e^{-1/2}-\frac{1}{3}e^{-3/2})\frac{\pi b}{\sqrt{2\lambda}}$ for $0\leq\epsilon\leq1$, where $e=2.71828\cdots$ is base of natural logarithm.

It is worth noting that the solution of BPS limit \cite{1} is not exact solutions for the metric around a tensor monopole, since it is not the solution of Eq.(5). In other words, Seifert's result \cite{1} only describes the scene of far-field. Eq.(12) is an approximative solution of the Kalb-Ramond field in the thin-wall limit. Therefore, the simplified model shares some features of the realistic solution for the coupled Einstein-Kalb-Ramond system of equations (5), (9) and (10), as we shall rigorously confirm in the next section.\\

\section{THE SOLUTION FOR THE COUPLED SYSTEM}
The asymptotic form of the functions $E(r)$, $F(r)$ and $g(r)$ can be systematically constructed in both regions, near the origin and for $r\rightarrow\infty$. Expanding the functions around the origin gives:
\begin{eqnarray}
E(r)&=&e_0(1+(\frac{3\epsilon}{b^2}g_1^2-\frac{b^2\epsilon\lambda}{12})r^2 \nonumber\\
&+&[\frac{27\epsilon^2}{10b^4}g_1^4+(\frac{7\epsilon\lambda}{10}+\frac{3\epsilon^2\lambda}{40})g_1^2]r^4+O(r^6))\\
F(r)&=&1+(\frac{3\epsilon}{2b^2}g_1^2+\frac{b^2\epsilon\lambda}{12})r^2 \nonumber\\
&+&[(\frac{13\epsilon^2\lambda}{40}-\frac{4\epsilon\lambda}{5})g_1^2-\frac{9\epsilon^2}{20b^4}g_1^4+\frac{b^4\epsilon^2\lambda^2}{144}]r^4+O(r^6)\nonumber \\ \\
g(r)&=&g_1(r+[(\frac{\epsilon}{20}-\frac{1}{5})\lambda b^2-\frac{9\epsilon}{20b^2}g_1^2]r^3 \nonumber\\  &+&[(\frac{1}{5}-\frac{13\epsilon}{60}+\frac{\epsilon^2}{24})\frac{\lambda^2b^4}{14}+(1-\frac{3\epsilon}{10}-\frac{3\epsilon^2}{20})\frac{\lambda g_1^2}{7}\nonumber \\&+&\frac{27\epsilon^2g_1^4}{160b^4}]r^5+O(r^7))
\end{eqnarray}
where $g_1$ and $e_0\equiv E(0)$ are free parameters to be determined numerically. The asymptotic behavior ($r\gg(\sqrt{2\lambda}b)^{-1}$) is given by
\begin{eqnarray}
E(r)&=&(\sqrt{2\lambda}br)^{2\epsilon}(1+\frac{(1-\epsilon)\epsilon}{4\lambda b^2(1+\epsilon)}\frac{1}{r^2}\nonumber\\
&-&\frac{\epsilon(1-\epsilon)^3}{8\lambda^2 b^4(3-\epsilon)(1+\epsilon)^2}\frac{1}{r^4}+O(\frac{1}{r^6})),\\
F(r)&=&(1+\epsilon)(1-\frac{\epsilon(1-\epsilon)}{4\lambda b^2(1+\epsilon)}\frac{1}{r^2}\nonumber\\
&+&\frac{\epsilon(\epsilon-1)(\epsilon^3-4\epsilon^2-5\epsilon+16)}{16\lambda^2b^4(3-\epsilon)(1+\epsilon)^2}\frac{1}{r^4}+O(\frac{1}{r^6})),\\
g(r)&=&\frac{b}{\sqrt{2}}-\frac{1-\epsilon}{2\sqrt{2}\lambda b(1+\epsilon)}\frac{1}{r^2}\nonumber\\
&-&\frac{(1-\epsilon)(3-\epsilon^2)}{8\sqrt{2}\lambda^2b^3(1+\epsilon)^2}\frac{1}{r^4}+O(\frac{1}{r^6}).
\end{eqnarray}
It is obviously that $F(r)$ will converge to $(1+\epsilon)$, but $E(r)$ grows without bound as $r\rightarrow \infty$ and $E(r)\propto r^{2\epsilon}$. From Eqs.(6)-(8), we have
\begin{eqnarray}
\rho+p_r+2p_\theta=\frac{4b^2}{r^2}+\frac{(1-\epsilon)^3}{(1+\epsilon)^2\lambda}\frac{1}{r^4}+O(\frac{1}{r^6})
\end{eqnarray}
which is proportional to the $tt$ component of the trace-reversed energy-momentum tensor and couples to the $tt$ component of the metric in the linearized approximation \cite{1}. On the contrary, $\rho+p_r+2p_\theta$ falls off as $r^{-4}$ for the $O(3)$ scalar monopole. Therefore, their gravitational fields have essential differences. If the mass scale $b$ is well below the Planck scale, the far-field shall become sufficiently flat so that the solution of tensor monopole can be embedded in one describing the suitable large-scale structure.

The limit of flat space is recovered for $\epsilon=0$, $e_0=1$ and $E(r)=F(r)=1$ in Eqs.(17)-(22), and $g_1$ is determined numerically. We do that through a fourth-order Runge-Kutta method for the quantity $\tilde{g}(\tilde{r})$, where $\tilde{g}\equiv\frac{g}{b}$ and $\tilde{r}=\sqrt{2\lambda}br$ is a dimensionless parameter. We impose the initial conditions at the origin $\tilde{g}(0)=0$ and $\dot{\tilde{g}}(0)=\frac{g_1}{\sqrt{2\lambda}b^2}$, where overdot denotes differentiation with respect to $\tilde{r}$. $g_1$ is adjusted so that $\tilde{g}\rightarrow\frac{1}{\sqrt{2}}$ for large $\tilde{r}$ using shooting routine. We display $\tilde{g}(\tilde{r})$ in Fig.1 for the case of flat spacetime. Next, we present the numerical solutions of the full system of field equations coupled to gravity. These solutions are the gravitating generalization of the flat spacetime one. To evaluate the solutions of full system by numerical method, the boundary conditions (11) can be reduced to
\begin{eqnarray}
&\tilde{g}&(0)=0,F(0)=1,E(0)=e_0,\noindent \nonumber\\
&\tilde{g}&(\tilde{R})=\frac{1}{\sqrt{2}},E(\tilde{R})=\tilde{R}^{2\epsilon},F(\tilde{R})=1+\epsilon,
\end{eqnarray}
up to $O(\tilde{R}^{-2})$ order for large $\tilde{R}$, where $\tilde{R}=\sqrt{2\lambda}bR$. To use shooting method, we impose the initial conditions at the origin $\tilde{g}(0)=0$, $\tilde{E}(0)=0$, $\tilde{F}(0)=0$ and $\dot{\tilde{F}}(0)=1$. The values of $\dot{\tilde{g}}(0)$ and $\dot{\tilde{E}}(0)$ are adjusted so that $\tilde{g}(\tilde{R})=\frac{1}{\sqrt{2}}$ and $\tilde{E}(\tilde{R})=\tilde{R}^{1-2\epsilon}$, where $\tilde{E}\equiv\frac{\tilde{r}}{E}$ and $\tilde{F}\equiv\frac{\tilde{r}}{F}$. In Fig.1, we display $\tilde{g}(\tilde{r})$ for $\epsilon=0$ and $10^{-2}$. The profile of $\tilde{g}(\tilde{r})$ is insensitive to $\epsilon$ for $0\leq\epsilon\leq1$ not only asymptotically, but also close to the origin. In Fig.2, the components of metric $E(\tilde{r})$ and $F(\tilde{r})$ are plotted vs the dimensionless coordinate $\tilde{r}=\sqrt{2\lambda}br$ for different $\epsilon$. Moreover, both $E(\tilde{r})$ and $F(\tilde{r})$ increase with $\epsilon$ increasing.
\begin{figure}
\includegraphics[height=2.5in,width=3.1in]{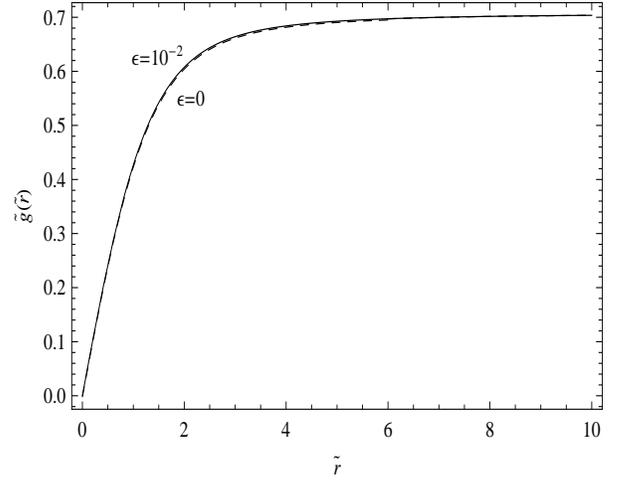}
\caption{The function of $\tilde{g}(\tilde{r})$ corresponding to the monopole configuration, is plotted vs the dimensionless coordinate $\tilde{r}=\sqrt{2\lambda}br$ for $\epsilon=0$ (dash line) and $10^{-2}$ (solid line). The shape of the curve is quite insensitive to the value $\epsilon$ in the range of $0\leq\epsilon\leq1$.}
\end{figure}\\
\begin{figure}
\includegraphics[height=2.5in,width=3in]{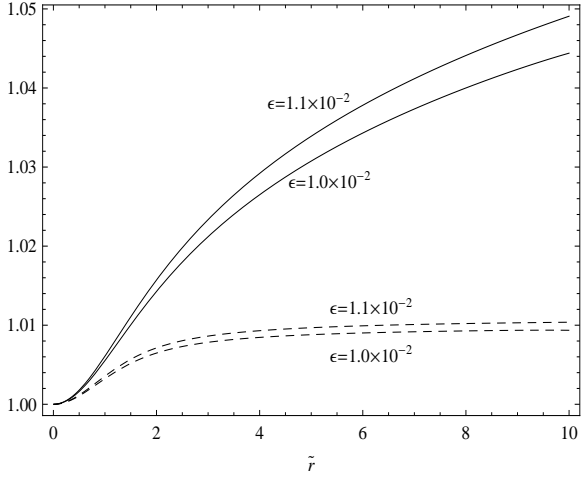}
\caption{$E(\tilde{r})$ (solid line) and $F(\tilde{r})$ (dash line) are plotted vs the dimensionless coordinate $\tilde{r}=\sqrt{2\lambda}br$ for different values of $\epsilon$. It is clearly seen from this figure that both $E(\tilde{r})$ and $F(\tilde{r})$ increase with $\epsilon$ increasing.}
\end{figure}\\

\section{OBSERVATIONAL SIGNATURES}
We now study the motion of test photons around a tensor monopole. Eqs.(20)-(22) are good approximation unless we were interested in the test photons moving right into the monopole core $\delta\sim(\sqrt{2\lambda}b)^{-1}$. In the case of BPS limit, Seifert have pointed out that the gravitational redshift experienced by a photon in the background of tensor monopole is within no more than $\epsilon^2$ order if the mass scale $b$ is well below the Planck scale \cite{2}. It is still kept that redshift effect will be negligible for the realistic solution of coupled system. However, the effect for the deflection of light by the gravitational field is more interesting. A null geodesics equation in the plane $\theta=\pi/2$ reads
\begin{equation}
-E(r)\dot{t}^2+F(r)\dot{r}^2+r^2\dot{\varphi}^2=0
\end{equation}
where dot denotes the derivative with respect to some affine parameter on the worldline. Since the metric is spherically symmetric and static, there are two Killing vector field $t^\mu$ and $\varphi^\mu$ leading to two constants of the motion: $\mathcal{E}=E^2\dot{t}$ and $\mathcal{J}=r^2\dot{\varphi}$. From Eq.(25), we have
\begin{equation}
\frac{d\varphi}{dr}=\pm\frac{1}{r^2}\frac{1}{\sqrt{\beta^{-2}\frac{1}{EF}-\frac{1}{Fr^2}}}
\end{equation}
where $\beta=\mathcal{J}/\mathcal{E}$. If $\mathcal{E}^2<(2+\frac{2}{3}\epsilon-\frac{7}{9}\epsilon^2+\frac{14}{9}\epsilon^3)\lambda b^2\mathcal{J}^2$,
we have $r_m>(\sqrt{2\lambda}b)^{-1}$, where $r_m$ is the value of $r$ for which the denominator of Eq.(26) vanishes. In other words, $r_m$ is the largest root of the equation $\beta^2E(r_m)=r_m^2$ and is larger than the core radius of the monopole. The orbit of the light ray will have a "turning point" at $r=r_m$. In this case, we have approximate expression of the total angular deflection up to $\tilde{r}_m^{-4}$  ($\tilde{r}_m=\sqrt{2\lambda}br_m$) order
\begin{eqnarray}
\Delta\varphi&=&\frac{\sqrt{1+\epsilon}}{1-\epsilon}(\pi-2Arcsin[\sqrt{\frac{\epsilon(1-\epsilon)}{2(1+\epsilon)}}\frac{1}{\tilde{r}_m}\nonumber\\&-&
(\frac{\sqrt{\epsilon}(1-\epsilon)^{5/2}}{2\sqrt{2}(3-\epsilon)(1+\epsilon)^{3/2}}+\frac{\epsilon^{3/2}(1-\epsilon)^{3/2}}{4\sqrt{2}(1+\epsilon)^{3/2}})\frac{1}{\tilde{r}_m^3}])
 \nonumber\\
&+&\frac{1}{2}\frac{\epsilon(1-\epsilon)}{\sqrt{1+\epsilon}(5-3\epsilon)}\frac{1}{\tilde{r}_m^2}-\frac{1}{2}\frac{\epsilon(1-\epsilon)}{(1+\epsilon)^{3/2}}\nonumber\\ &\times&[\frac{3-\epsilon^2}{5-\epsilon}+\frac{11-7\epsilon-\epsilon^2+\epsilon^3}{2(3-\epsilon)(7-3\epsilon)}+\frac{3\epsilon(1-\epsilon)}{4(5-3\epsilon)}]\frac{1}{\tilde{r}_m^4}.\nonumber\\
\end{eqnarray}
Defining $\delta\varphi\equiv\Delta\varphi-\pi$ to be the angle between the "unperturbed" and "perturbed" directions of propagation up to $\epsilon$ order
\begin{equation}
\delta\varphi\approx\frac{3}{2}\pi\epsilon-\sqrt{2\epsilon}\tilde{r}_m^{-1}+\frac{\epsilon}{10}\tilde{r}_m^{-2}+\frac{\sqrt{2\epsilon}}{6}\tilde{r}_m^{-3}-\frac{181\epsilon}{420}\tilde{r}_m^{-4}
\end{equation}
Obviously, $\delta\varphi\approx\frac{3}{2}\pi\epsilon$ in the case of $\tilde{r}_m\gg\epsilon^{-1}$, i.e, we repeat Seifert's approximation \cite{2}. By using same techniques, we obtain the angular deflection $\delta\varphi_s$ for the case of $O(3)$ scalar monopole \cite{15}
\begin{equation}
\delta\varphi_s\approx\frac{\pi}{4}\epsilon_s-\sqrt{3\epsilon_s/2}\tilde{r}_m^{-1/2}+\frac{25\epsilon_s}{96}\tilde{r}_m^{-2}+\sqrt{2\epsilon_s/3}\tilde{r}_m^{-5/2}-\frac{\epsilon_s}{8}\tilde{r}_m^{-4}
\end{equation}
where $\epsilon_s=16\pi G\eta^2$ and $\eta$ is mass scale in the $O(3)$ scalar monopole. For contrasting the two species of monopoles, we take $\eta=\sqrt{3/2}b$, Eq.(29) can be rewritten as
\begin{equation}
\delta\varphi_s\approx\frac{3}{2}\pi\epsilon-3\sqrt{\epsilon}\tilde{r}_m^{-1/2}+\frac{25\epsilon}{16}\tilde{r}_m^{-2}+2\sqrt{\epsilon}\tilde{r}_m^{-5/2}-\frac{3\epsilon}{4}\tilde{r}_m^{-4}
\end{equation}
Therefore, the deflection angles have important qualitative differences between the tensor and scalar monopoles. It furnishes a possibility that two species are discriminated by the observation of light rays in these backgrounds. In Fig.3, we plotted the $\delta\varphi$ and $\delta\varphi_s$ vs the parameter $r_m$ for a typical grand unification scale $b\sim10^{16}GeV$. By the numerical calculation, we show that $\delta\varphi$ is quite insensitive and $\delta\varphi_s$ is sensitive in the same interval of $r_m$. In Fig.4, we plotted the $\delta\varphi$ and $\delta\varphi_s$ vs the apparent impact parameter $\beta$ for $\epsilon=10^{-2}$. Both tensor and scalar monopoles have tiny core radius $\delta\sim\delta_s\sim(\sqrt{2\lambda}b)^{-1}$, therefore Eqs.(28) and (30) are very accurate expressions when $r_m$ is far larger than the core radius. Set $\theta_{max}$ is the maximum of $|\delta\varphi_s-\delta\varphi|$, we have $\theta_{max}\sim\sqrt{\epsilon\delta/r_m}$. If the source of light, the monopole, and the observer are aligned exactly, all the rays that pass at the appropriate parameter $r_m$ around the monopole, at any azimuth, reach the position of the observer. Under this special circumstance, the observer sees an infinite number of images, which form a ring around the monopole. Assuming the source is much farther from the monopole, its rays are then nearly parallel to the line of alignment, and the deflection angle required for the ray to reach the observer is $r_m/D$, where $D$ is the distance from the monopole to the observer. Thus, the angular radius of Einstein ring is $r_m/D\approx\frac{3}{2}\pi\epsilon$ unless a monopole is nearing the solar system, which leads to $\theta_{max}\ll10^{-9}$ radians. For $\epsilon=10^{-2}$ and $D\approx10^4$ light-years, we have $r_m/D\approx0.05$ radians and $\theta_{max}\approx10^{-25}$ radians. By means of observation of Einstein ring, a monopole is able to find but it is powerless to determine whether or not to correspond to a tensor one. Once that the Einstein rings are discovered, we have to go a step further by the deflection of light near the monopole.

For a light ray just grazing the monopole, the effect is quite evident. If $r_m\sim10^{18}\epsilon\delta$, we have $\theta_{max}\sim10^{-9}$ radians. This angle is at the limit of resolution of telescope at present, so it can be observed when $r_m<10^{18}\epsilon\delta$. On the other hand, the star near the Sun are visible only during a total eclipse of the Sun, and even then brightness of the solar corona restricts observations to $r_m>2R_\odot$. In the case of monopole, there do not exist these difficulties since the monopole is a cold and dark object.

In conclusion, we have found that the deflection angles have important qualitative differences between tensor and scalar monopoles. This phenomenon might open up new direction in the search of Lorentz violation. If a monopole were detected, it would be discriminated whether or not corresponding to tensor one. Furthermore, tensor monopole would provide insight into the roles played by Lorentz symmetry in physics.
\begin{figure}
\includegraphics[height=2.5in,width=3in]{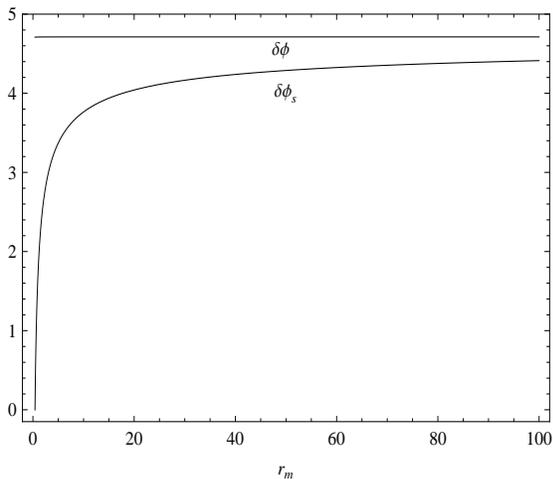}
\caption{The functions $\delta\varphi$ and $\delta\varphi_s$ are plotted for $\epsilon=16\pi Gb^2=10^{-6}$, where the unit of $r_m$ is $\frac{1}{\sqrt{2\lambda}b\epsilon}$ and unit of deflection angle is $\epsilon$. The shape of the $\delta\varphi$ curve is quite insensitive to the value of $r_m$ in the interval $\frac{1}{\sqrt{8\lambda}b\epsilon}\leq r_m\leq\frac{25}{\sqrt{8\lambda}b\epsilon}$. On the contrary, the shape of $\delta\varphi_s$ is sensitive in this interval.}
\end{figure}
\begin{figure}
\includegraphics[height=2.5in,width=3in]{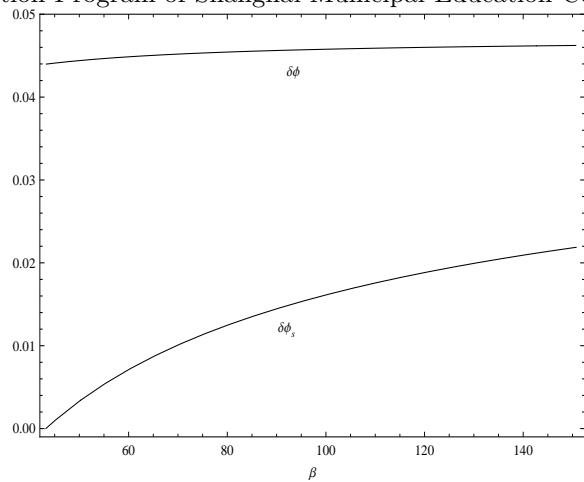}
\caption{The variations of $\delta\varphi$ and $\delta\varphi_s$ with $\beta$ are plotted for $\epsilon=16\pi Gb^2=0.01$, where the unit of $\beta$ is $\frac{1}{\sqrt{2\lambda}b}$.The shape of the $\delta\varphi_s$ curve is more sensitive to $\beta$ than that of $\delta\varphi$, although both curves tend to $\frac{3}{2}\pi\epsilon$ when $\beta$ is large enough.}
\end{figure}
\begin{acknowledgments}
This work is supported by National Education Foundation of China under grant No. 200931271104 and Innovation Program of Shanghai Municipal Education Commission (12YZ089).
\end{acknowledgments}

\end{document}